\documentclass[lettersize,journal]{IEEEtran}
\usepackage{amsmath,amsfonts,color}
\usepackage{algorithmic}
\usepackage{algorithm}
\usepackage{array}
\usepackage{textcomp}
\usepackage{stfloats}
\usepackage{url}
\usepackage{verbatim}
\usepackage{graphicx}
\usepackage{cite}
\usepackage{epsfig,epstopdf}
\usepackage{subfigure}
\usepackage{makecell}
\usepackage{multirow}
\usepackage{diagbox}
\usepackage{stfloats}
\usepackage{float}
\graphicspath{{figure/}}
\newtheorem{remark}{Remark}
\begin{document}
\title{Lateral Control of Brain-Controlled Vehicle Based on SVM Probability Output Model}

\author{Hongguang Pan, \emph{Member,~IEEE}, Xinyu Yu, Yong Yang}
\maketitle
\thanks{This work was jointly supported by the National Natural Science Foundation of China under Grant No.U1964202, 61603295, Qin Chuangyuan ``Scientists + Engineers'' Team Construction in Shaanxi Province under Grant No. 2022KXJ-38, and Xi'an Science and Technology Program under Grant No.2022JH-RGZN-0041.

Hongguang Pan, Xinyu Yu and Yong Yang are with College of Electrical and Control Engineering, Xi'an University of Science and Technology, Xi'an 710054, China, and Xi'an Key Laboratory of Electrical Equipment Condition Monitoring and Power Supply Security. Xi'an 710054 (e-mail: hongguangpan@163.com, 15035804802@163.com, yongy@xust.edu.cn, llliuzesheng@163.com)}

\begin{abstract}
The non-stationary characteristics of EEG signal and the individual differences of brain-computer interfaces (BCIs) lead to poor performance in the control process of the brain-controlled vehicles (BCVs). In this paper, by combining steady-state visual evoked potential (SSVEP) interactive interface, brain instructions generation module and vehicle lateral control module, a probabilistic output model based on support vector machine (SVM) is proposed for BCV lateral control to improve the driving performance. Firstly, a filter bank common spatial pattern (FBCSP) algorithm is introduced into the brain instructions generation module, which can improve the off-line decoding performance. Secondly, a sigmod-fitting SVM (SF-SVM) is trained based on the sigmod-fitting method and the lateral control module is developed, which can produce all commands in the form of probability instead of specific single command. Finally, a pre-experiment and two road-keeping experiments are conducted. In the pre-experiment, the experiment results show that, the average highest off-line accuracy among subjects is 95.64\%, while for those in the online stage, the average accuracy is only 84.44\%. In the road-keeping experiments, the task completion rate in the two designed scenes increased by 25.6\% and 20\%, respectively.
\end{abstract}

\begin{IEEEkeywords}
Brain-controlled vehicle, lateral control, filter bank common spatial pattern, sigmod-fitting method, probability output model.
\end{IEEEkeywords}
\section{Introduction}
\IEEEPARstart{B}{rain}-controlled vehicle (BCV) is a vehicle controlled by the human brain rather than limbs through brain-computer interface, which can convert Electroencephalogram (EEG) signals into control instructions and establish direct control channels between human intentions and external devices\cite{River2022,Fish2016,Ola2016,Teo2022}. On the one hand, for the disabled persons, BCVs may help them recover their ability to drive, expand their range of activities, and improve their quality of life. On the other hand, for the healthy individuals, BCV can liberate their limbs and enhance their driving experience by providing a new driving style\cite{wang2022}.

EEG signal has become the most commonly used technology in the brain-computer interface (BCI) system because of its high time resolution, high signal-to-noise ratio (SNR) and easy access. At present, BCI system based on EEG signal is used to control robots directly in most kinds of work through users' minds\cite{pan2022,li2022,xu2022,Xin2022,Beraldo2022}. It has significant advantages in rehabilitation fields, such as rehabilitation for the disabled and nursing care for the elderly\cite{xiao2022,Guo2022}. Steady-state visual evoked potential (SSVEP) is a periodic nerve response induced by repeated visual stimuli, which usually generates EEG signals with the same frequency modulation as the target. In this paper, we choose SSVEP as an experimental paradigm, because it is suitable for the steering system of BCVs\cite{zheng2014,Lu2019}.

Bi \emph{et al}.\cite{zheng2014} used SSVEP-BCI based on HUD for lateral control of vehicle. In the experiment part, the subjects completed the task on the U-shaped road. The experimental results verify the possibility of using EEG to carry out vehicle lateral control at low speed. However, due to the degradation of the online BCI system performance, some subjects overall experimental results are not very ideal.  Lu and Bi proposed an EEG-based longitudinal control system for brain-controlled vehicles\cite{Lu2019}. This method is tested in the virtual scene of the laboratory by simulating experiments on vehicles. The average accuracy of the off-line BCI system is more than 90\%. But it shows relatively poor driving performance for some subjects in the online stage, and the longitudinal control module can only output three fixed commands, considerably reducing the performance of the BCV. Given the constraints of the limited of all existing BCI systems, it is very important to finding ways to enhance and ensure the overall driving performance.

\begin{figure*}[t]
\subfigure[]{
\begin{minipage}{0.66\linewidth}
\centering%
\includegraphics[height=4.5cm]{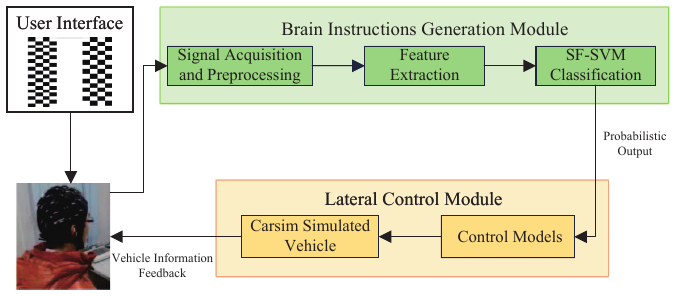} \label{fig2-1a}
\end{minipage}}
\subfigure[]{
\begin{minipage}{0.33\linewidth}
\centering%
\includegraphics[height=4.5cm]{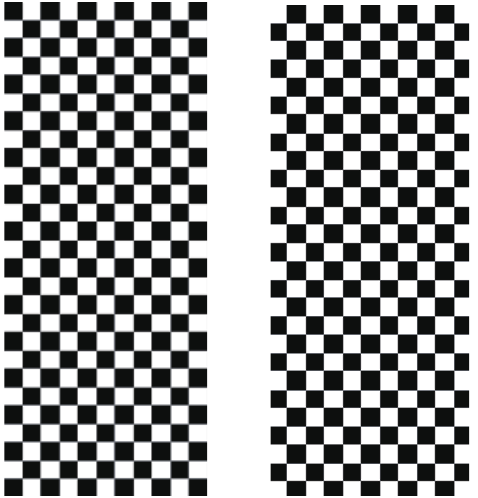}\label{fig2-1b}
\label{fig:side:b}
\end{minipage}}
\caption{Vehicle lateral control system: (a) the structure of lateral control system, (b) user Interface: checkerboard pattern and its reversed form.}
\end{figure*}

At present, some work used auxiliary controllers and shared control strategies \cite{yan2023,Ans2022,Hang2021} to improve the BCVs driving. Zhu \emph{et al}.\cite{zhuang2021} established a BCI system using the motion imagination to control the simulated vehicle using a shared control strategy. Lu \emph{et al}.\cite{laYun2020} used model predictive control (MPC) as an auxiliary controller to control the lateral direction of the vehicle. The simulation results showed that the method makes it possible to test road-keeping and avoid obstacles while maintaining the user's control authority. However, as we all know, introducing MPC as an auxiliary controller will change the control system structure and bring a lot of computation. In addition to the above methods, Khan \emph{et al}.\cite{khan2021} designed a brain-driven intelligent wheelchair based on a vision-based sensor network, which can be controlled by SSVEP brain signal. In off-line training stage, the accuracy was up to 96\%, meanwhile the 95\% accuracy was achieved for online control.

In this paper, we introduced FBCSP algorithm into the brain instructions generation module, which can improve the off-line decoding performance. On this basis, lateral control module based on the SF-SVM method was developed to improve driving performance. The contributions of this paper consists of the following three parts:
\begin{enumerate}
  \item Improve the SNR of the EEG signals and the off-line decoding performance of the brain instructions generation module, through designing and introducing the filter bank common spatial pattern (FBCSP) algorithm.
  \item Improve the driving performance of BCV, sigmod-fitting SVM (SF-SVM) is trained based on the sigmod-fitting method. Based on the SF-SVM, the vehicle lateral control module is designed, this module can convert output instructions into probability values.
  \item Verify the effectiveness of the proposed overall scheme. In the pre-experiment, the experiment results show that the average highest off-line accuracy among subjects is 95.64\%, the average online accuracy is only 84.44\%. In the road-keeping test, the lateral control module based on the SF-SVM has better BCV driving performance.
\end{enumerate}


\section{Lateral Control System Design} \label{sec2}
As shown in Fig.\ref{fig2-1a}, the vehicle lateral control system proposed in this paper is made up of two parts, i.e., brain instructions generation module and lateral control module. The working process is as follows. The driver makes a decision based on the feedback from the surroundings to generate a steering command. At the same time, the driver needs to look at the corresponding stimulus on the user interaction interface, and then decodes the collected EEG signal into a control command. Finally, the lateral control module converts the command into the corresponding control signal and transmits it to control the vehicle.
\subsection{Interaction Interface Design}\label{sec2.1}
This study is based on brain vision, and uses the steady state motion reversal visual stimulation paradigm with low flicker and the tireless user to induce periodic steady state potentials. It mainly consists of two chessboards, with the reversal frequency of 13Hz and 11Hz, respectively. Each chessboard is coded as 11$\times$25 cells (white and black), each size is 25 pixels. The reasons for choosing the frequency of this paradigm are as follows. Firstly, low-frequency stimulation is not easy to cause brain fatigue. Then, the spectral peak generated by the SSVEP is around 10 Hz, and the amplitude reaches the maximum at 15 Hz and 12Hz\cite{Pastor2003}. Fig.\ref{fig2-1b} presents the checkerboard pattern.

The left and right chessboards are used for lateral control of the vehicle. When users want to turn right, they need to look at the right chessboard, inducing the brain to generate EEG signals of the corresponding frequency. When they want to turn left, they need to watch the left chessboard. When the user wants to maintain the current driving state, there is no need to look at any stimulus.
\subsection{Brain Instructions Generation Module}\label{sec2.2}
\begin{figure}[h]
\centering%
\includegraphics[scale=0.58]{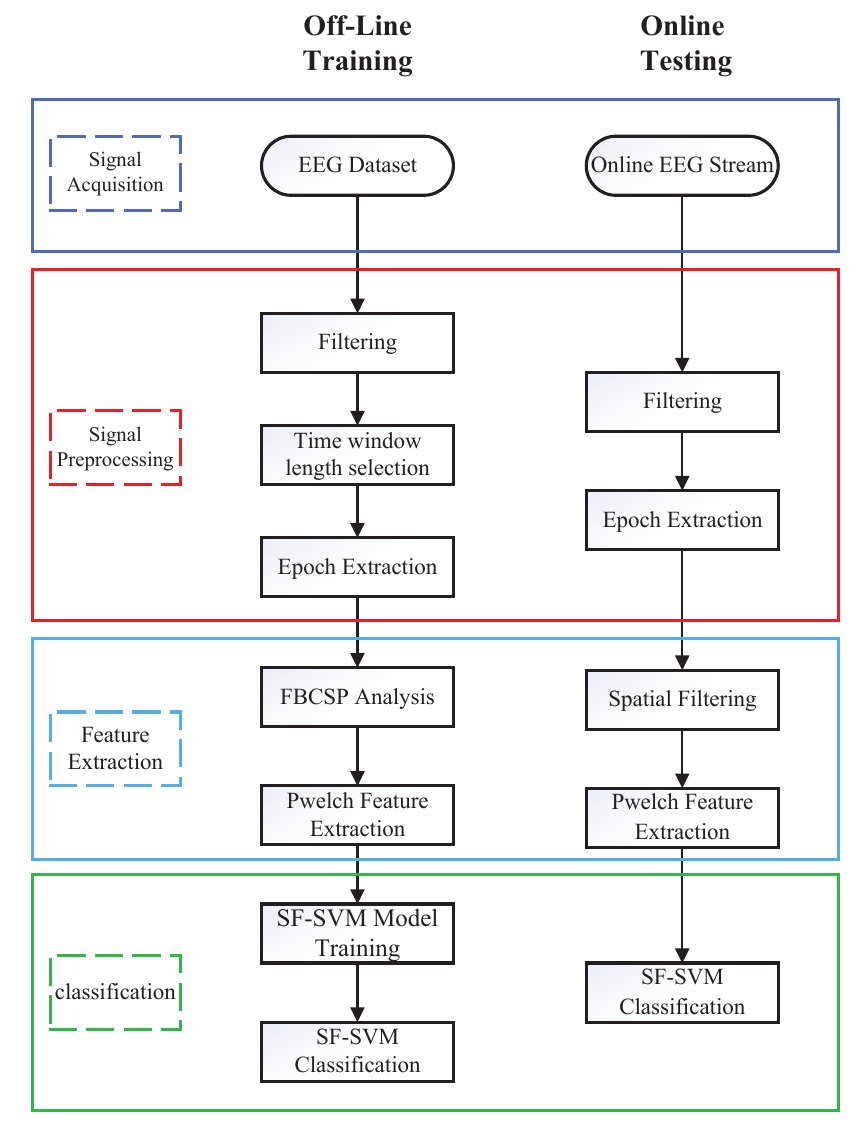}
\caption{The structure of brain instructions generation module.}  \label{fig2-3}
\end{figure}
The signal flow diagram of the brain instructions generation module is shown in Fig.\ref{fig2-3}, which consists of off-line and online stages. The off-line phase is used to decode the EEG signal, and the key parameters and models are trained for subsequent use in the online stage. Key parameters include the length of the time window, the projection matrix generated by the FBCSP, and the key parameters in the SF-SVM model. In the online phase, EEG is collected in real time, and the signal is decoded by off-line model to generate the corresponding instructions. These two parts are mainly divided into the following three steps: signal acquisition and preprocessing, feature extraction, and classification.

\subsubsection{Signal Acquisition and Preprocessing}\label{sec2.2.1}
\begin{figure}[h]
\centering%
\includegraphics[scale=0.52]{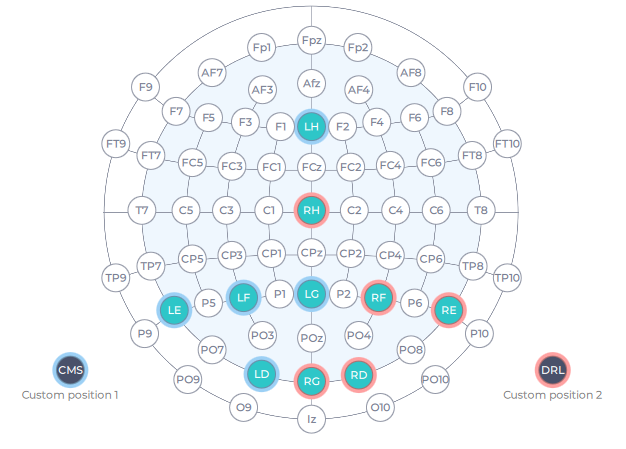}
\caption{The location of the EEG electrode used.}\label{fig2-4}
\end{figure}

The signal acquisition equipment is the EPOC Flex-32 Channel Wireless EEG Headset of Emotiv company, and the electrode placement position is Cz, Fz, O1, O2, Oz, P3, P4, P7, P8, and Pz. The electrode position is shown in Fig.\ref{fig2-4}. The reference potential corresponds to the average potential of the left and right earlobes. The sampling frequency is 128 Hz, the contact impedance between the electrode and scalp is calibrated to be less than 10 k\emph{$\Omega$}. The collected signal needs to be band-pass filtered (4-49Hz).

EEG signal is converted into a corresponding instruction for each time window length. The length of time windows is one of the key parameters, which determines the decoding performance. Therefore, in the training phase, this study tests the decoding performance by defining different time window lengths to ensure that the brain instructions generation module has the best decoding performance. It should be pointed out that in order to increase the training dataset, we set the step size of the time window to 0.5s at the training stage.

\begin{figure}[tp]
\centering%
\includegraphics[scale=0.32]{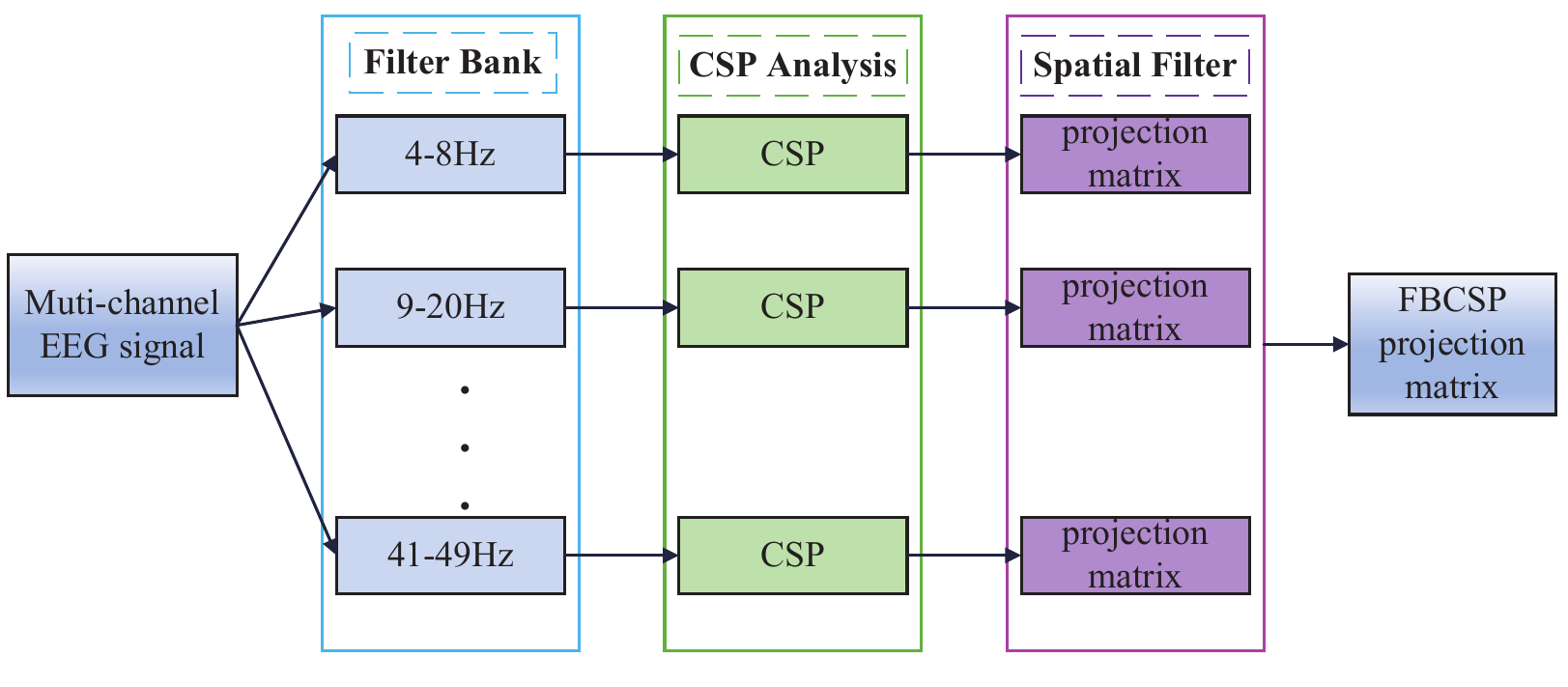}
\caption{The structure of filter bank common spatial pattern.}\label{fig2-5}
\end{figure}

\subsubsection{Feature Extraction}\label{sec2.2.2}
The spatial filter is used to improve the resolution of multi-channel EEG signals and the SNR of EEG signals. In this section, in order to take into account the interaction between different sub-bands of EEG signals, we apply one-versus-rest FBCSP algorithm to generate spatial filters. The overall algorithm of the FBCSP is shown in Fig.\ref{fig2-5}.

For two classification tasks, a multi-channel EEG epoch $E$ with dimensions $\textrm{N} \times \textrm{S}$, where $\textrm{N}$ is the number of channels, and $\textrm{S}$ is the sample points number of EEG epoch per channel, the normalized covariance matrix $R_c$ can be expressed as:
\begin{align}
R_c=\frac{E_{c} E_{c}{ }^\textrm{T}}{\textrm{trace}\left(E_{c} E_{c}{ }^\textrm{T}\right)}, c \in\{1,2\} \label{eq1}
\end{align}
where $E_{c}^\textrm{T}$ represents the transposition of matrix $E_{c}$, $\textrm{trace}()$ represents the summation of elements on the diagonal of the matrix. Then covariance matrix $R$ of mixed spaces can be factorized as:
\begin{align}
R=\overline{R_1}+\overline{R_2} = U \lambda U^{\textrm{T}} \label{eq2}
\end{align}
where $\overline{R_1}$ and $\overline{R_2}$ are the average covariance matrix by averaging over all the trials of each classification task, $U$ is the matrix of eigenvectors and $\lambda$ is the diagonal matrix of eigenvalues. The whitening characteristic matrix $Z$ is shown in \eqref{eq3}:
\begin{align}
Z = \frac{1}{\sqrt{\lambda}} U^{\textrm{T}} \label{eq3}
\end{align}

The average covariance matrices$\overline{R_1}$,$\overline{R_2}$ can be transformed by $P$:
\begin{align}
S_{1}=Z \overline{R_{1}} Z^\textrm{T} \quad S_{2}=Z \overline{R_{2}} Z^\textrm{T} \label{eq4}
\end{align}
where $S_{1}$ and $S_{2}$ have same eigenvector $B$. By performing principal component decomposition for two matrices, it can be obtained that:
\begin{align}
S_{1}=B \lambda_{1} B^\textrm{T} \quad S_{2}=B \lambda_{2} B^\textrm{T} \quad \lambda_{1}+\lambda_{2}=I \label{eq5}
\end{align}

Therefore, the projection matrix $W$ is denoted as:
\begin{align}
W=B^\textrm{T} Z \label{eq6}
\end{align}

In this section, for multi-class problems, we apply one-versus-rest algorithm to generate spatial filters. Firstly, we filter the EEG signal of different frequency bands. Then, $W_{\textrm{ave},a}$ is obtained by taking the average of the projection matrix of each frequency band:
\begin{align}
W_{\textrm{ave},a}=\frac{\Sigma B^{\textrm{T}}_m Z_m}{M} , a \in\{1,2,3\}, m = 1\cdots M, \label{eq7}
\end{align}
where $m$ is the number of frequency sub-bands, $Z_m$ is whitening matrix of the $m$-th frequency band, $B^{\textrm{T}}_m$ is the feature vector of the $m$-th frequency band. $D_a \in \mathbb{R}^{\textrm{N} \times \textrm{S}}$ is the EEG signal obtained by filtering the $a$-th average projection matrix through spatial filtering:
\begin{align}
D_a=W_{\textrm{ave},a} E, a \in\{1,2,3\}\label{eq8}
\end{align}
where $W_{\textrm{ave},a} \in \mathbb{R}^{\textrm{N} \times \textrm{N}}$ is $a$-th CSP average projection matrix.

The frequency information of the EEG epoch is obtained by using the Pwelch algorithm which uses a short sliding window $I(t)$, continuing intercept the signal $x(t)$ in sections, and then perform fourier transform on the windowed intercepted signal:
\begin{align}
\operatorname{Pwelch}(t, \omega)=\int x(\tau) I(\tau-\mathrm{t}) e^{-j \omega t} d \tau \label{eq9}
\end{align}
where $\omega$ is the angular frequency, the Hanning window is selected as a window function $I(t)$ and the window function is:
\begin{align}
I(\mathrm{t})=\left\{\begin{array}{l}
0.5\left\{1-\cos \left(\frac{2 \pi t}{T}\right)\right\}, 0 \leq t \leq T \\ \label{eq10}
0, \text { other }
\end{array}\right.
\end{align}

And the expression for the signal power spectrum is:
\begin{align}
f(\omega, \mathrm{t})=|\operatorname{Pwelch}(\omega, t)|^2 / 2 \pi \label{eq11}
\end{align}

In this section, we extract the power spectrum of each channel as the characteristics. The power spectrum extraction for each channel of the EEG epoch at the specific frequency of half frequency $\pm$0.5Hz, basic frequency $\pm$0.5Hz, double frequency $\pm$0.5Hz. So we can get 60 features for each EEG epoch, the feature extraction is shown in \eqref{eq12}:
\begin{align}
f_a=\left[f_{a,1}, f_{a,2}, \ldots, f_{a,k}, \ldots, f_{a,60}\right]^\textrm{T}\label{eq12}
\end{align}
where $f_{a,k}$ denotes the $k$-th power spectrum feature of EEG epoch filtered by the $a$-th CSP projection matrix.

\subsubsection{Classification Strategy Based on SF-SVM}\label{sec2.2.3}
We adopted a one-versus-rest classification strategy to achieve multi-classification. The one-versus-rest classification strategy needs to build $n$ binary classifiers for the $n$ classification problem. Compared with traditional learning methods, the final decision function of support vector machine is only determined by a few support vectors, thus avoiding the disaster of dimensionality. In this paper, SVM with RBF as kernel function is used to construct classifier. For $i$-th EEG epoch, $y^a_i $ is $a$-{th} SVM output without threshold, which can be depicted in \eqref{eq13}:

\begin{align}
y^a_i=\sum_{k=1}^{K^a} w_i^{k,a} \exp \left(-g\left\|x_i^{k,a}-x_i\right\|^2\right)+b^a_i \label{eq13}
\end{align}
where $K^a $ denotes the number of the support vector of the $a$-th classifier. For $i$-th EEG epoch, $x_i^{k,a}$ is the $k$-th support vector of $a$-th classifier, $w_i^{k,a}$ is the weight of the $k$-th support vector of $a$-th classifier, and $b_i^a$ is the bias of the $a$-th classifier.

The sigmod-fitting algorithm proposed by Platt is used to train the one-versus-rest SVM probability output value model. This method transforms the output of the classification model into probability values based on the distribution of categories. The classification model output of the training set is filtered by the sigmoid function, the probability output value of the $a$-th SVM model is shown in \eqref{eq14}:
\begin{align}
P_i^a=\frac{1}{1+\exp \left(A_{a} y_i^a+B_{a}\right)} \label{eq14}
\end{align}
where $A_a, B_a$ are the coefficient matrix of probability value output model, which can be obtained by optimizing the loss function:
\begin{align}
\min \{-\sum_i [H_i^a \log \left(P_i^a\right)+\left(1-H_i^a\right) \log \left(1-P_i^a\right)]\} \label{eq15}
\end{align}
where $H_i^a \in {\{-1,1\}}$ is the $i$-th label of $a$-th classifier.

After sigmod-fitting, each EEG epoch can be express as three corresponding three probability values, and the maximum probability value will be selected as the decoding result of this EEG epoch. We identify that a probability value greater than 0.9 indicates that the classification result is correct.

\subsection{Lateral Control Module}\label{sec2.3}
In section \ref{sec2.2}, the sigmod-fitting method is used to obtain the SF-SVM classification model, which converts the output instructions into probability values. The $\delta^{(n)}$ is the steering wheel angle by $n$-th update can be represented as in \eqref{eq16}:
\begin{align}
  \delta^{(n)}= \begin{cases}\min \left\{\delta^{(n-1)}+\Delta\delta \times P^{(n)}, \delta_{\max }\right\}, &L^{(n)}=1, \\ \delta^{(n-1)}, & L^{(n)}=0, \\ \max \left\{\delta^{(n-1)}-\Delta\delta \times P^{(n)} , \delta_{\min }\right\}, & L^{(n)}=-1 .\end{cases} \label{eq16}
\end{align}
where $L^{(n)}$ is the steering command output by the brain instructions generation module at the $n$-th update, $P^{(n)}$ is the probability value obtained by the SF-SVM. $\delta_{\max}$ and $\delta _{\min}$ are the maximum and minimum value range of steering wheel angle, respectively. The unit of $\delta$ is deg.

\section{Pre-Expriment}\label{sec3}
In this section, the subjects are required to complete the pre-experiment so as to train and test the decoding performance of the brain instructions generation module. We select the subjects who performed both well in the off-line and online test because the BCVs cannot work for all users. At the same time, we consider that those well-behaved subjects were able to use brain signals to drive, thus allowing them to participate in the road-keeping experiment.

\begin{figure*}[t]
\subfigure[]{
\begin{minipage}{0.50\linewidth}
\centering%
\includegraphics[scale=0.55]{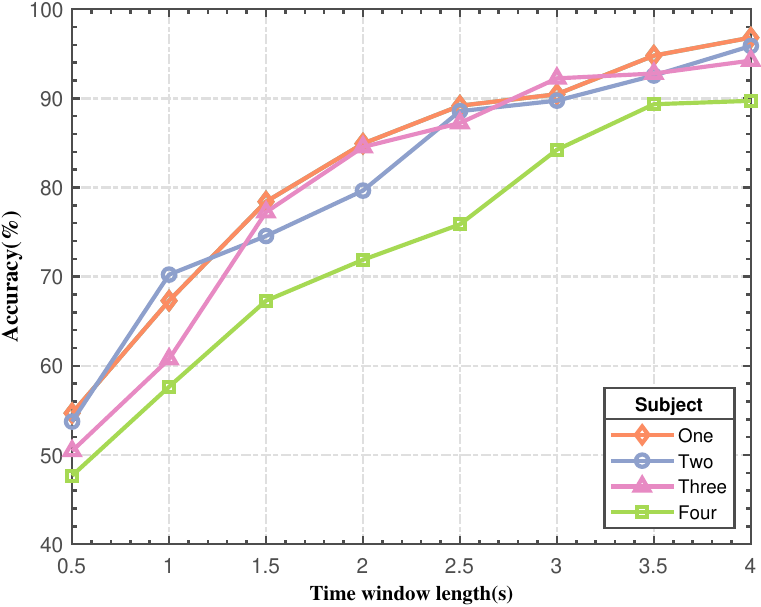} \label{fig3-2a}
\end{minipage}}
\subfigure[]{
\begin{minipage}{0.50\linewidth} \label{fig3-2b}
\centering%
\includegraphics[scale=0.55]{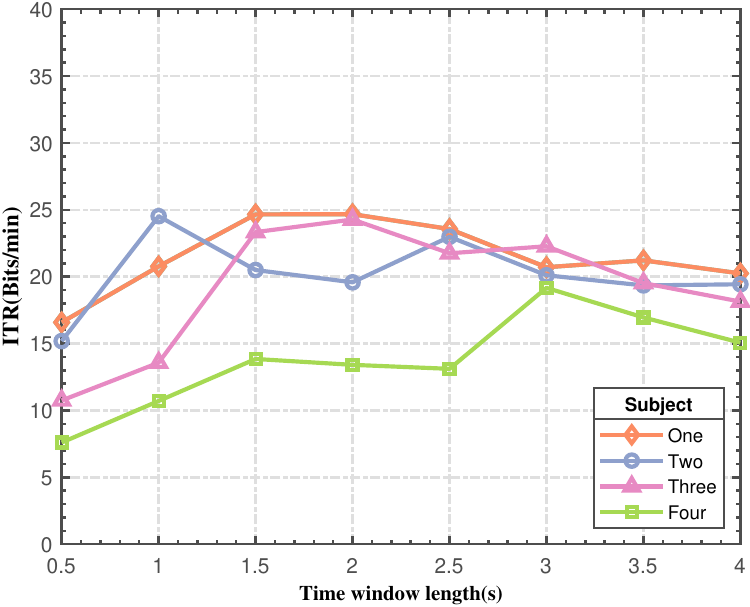}
\end{minipage}}
\caption{Static test results under different time window lengths of EEG epochs over subjects: (a) the cross-validations accuracy, (b) the ITR.}\label{fig3-2}
\end{figure*}

\begin{figure*}[t]
\subfigure[]{
\begin{minipage}{0.50\linewidth}
\centering%
\includegraphics[scale=0.55]{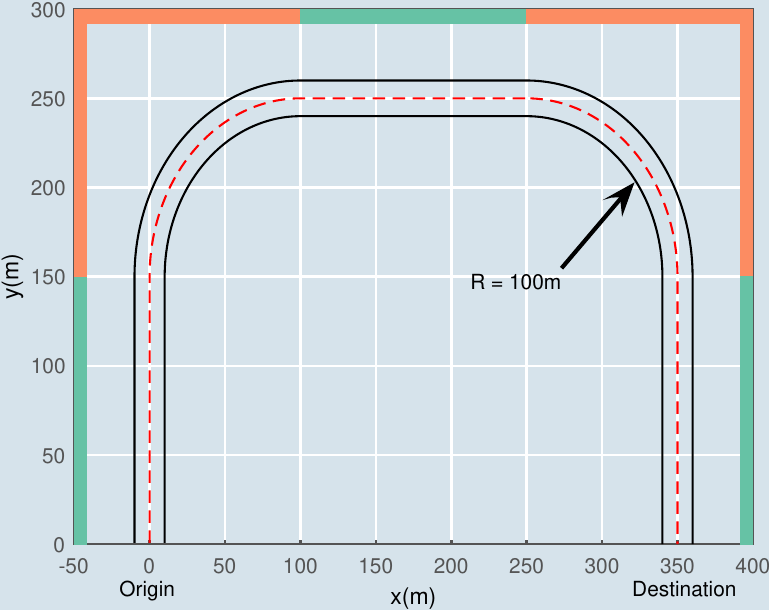}
\end{minipage}}
\subfigure[]{
\begin{minipage}{0.50\linewidth}
\centering%
\includegraphics[scale=0.55]{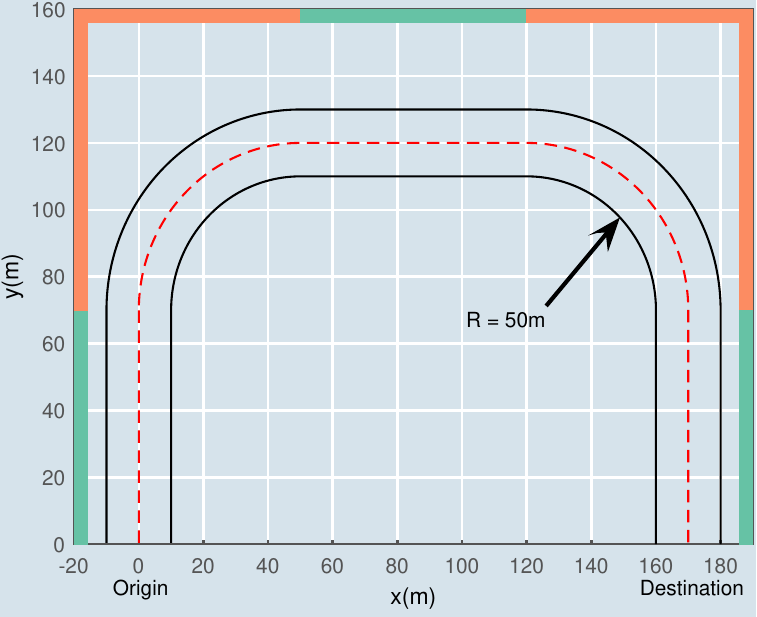}
\end{minipage}}
\caption{Road-keeping experiment scenes: (a) scene 1, (b) scene 2.}
\label{fig3-1}
\end{figure*}

\subsection{Experiment Subjects and Platform}\label{sec3.1}
Four subjects, aged from 22 to 25 years old (average age 23.8 years old), participate in this pre-experiment in this study. These subjects have no history of brain illness and have normal or adjusted vision.

The experimental platform is made up of EEG signal acquisition equipment, SSVEP stimulation interface, Carsim vehicle and lateral control module. The hardware environment of simulation experiment includes: the processor of Inter(R) Core(TM) i7-8750H, the graphics card of NVIDIA GeForce GTX 1050, and DDR4 16G of RAM. Emotiv's EPOC Flex-32 Channel Wireless EEG Headset is used to collect EEG signals and the SSVEP visual stimulation written in C Code. CarSim provides a simulation vehicle, and its lateral control module is built in the simulink of MATLAB. The communication system between the EEG signal acquisition equipment and Carsim is built by using Simulink.

\subsection{Experiment Process}\label{sec3.2}
The experiment includes the off-line and online stages. In the off-line stage, first, the subjects are asked to complete the EEG signals acquisition; then, the five-fold cross-validations for the accuracy of each subject is adopted to take the average results as the off-line accuracy. EEG acquisition is divided into the following two steps. First, participants are required to complete two groups of EEG signal acquisition, which are used to acquire the EEG signals at two frequencies. The two different frequencies represent the control commands for turning left and right respectively. For each frequency, the subjects are required to conduct four sessions, four trials for each session, each trial lasts for 12 seconds. Then, the subjects completed the test without any stimulation to acquire the EEG data related to the command of keeping straight driving. At this group, subjects need to conduct two sessions of collection, and each session carries out eight trials, each trial lasts for 12 seconds.

In the online stage,  the subjects, who perform well in the off-line stage, are also required to complete the online performance test. We select the time window length corresponding to the highest accuracy for off-line test, and the accuracy of three commands (turning right, turning left, gonging forward) will be tested, respectively.

\subsection{Experiment Results}\label{sec3.3}
The experiment results in the off-line stage are shown in Figs.\ref{fig3-2}, where Fig.\ref{fig3-2a} and Fig.\ref{fig3-2b} show the average accuracy and information translate rate (ITR) of the subjects under different time window lengths. For subject one, two and three, the average highest off-line accuracy is 95.64\%. Subject one exhibits the best decoding performance, and subject four presents the worst decoding performance who highest off-line accuracy is 89.72\%. \textbf{Therefore, subject four won't participate in the subsequent experiments.}

Table \ref{tab1} shows the average online accuracy of the other three subjects. This result is obtained under the time window length corresponding to the highest accuracy for each subject. Compared with their off-line performance, all three subjects show a certain decline in the performance of the online test and the average classification accuracy is only 84.44\%.
\begin{table}  \label{tab1}
\centering
\caption{Online decoding performance} \label{tab1}
\renewcommand\arraystretch{1.8}
\setlength{\tabcolsep}{1mm}{
\begin{tabular}  {cccccccc}
\hline \hline
Subject & \makecell[c]{Turning left \\accuracy(\%)} & \makecell[c]{Turning right\\accuracy(\%)} & \makecell[c]{Going Forward\\accuracy(\%)}& Mean accuracy(\%)\\
\hline S1 & 86.76 & 82.75 & 90.12 & 86.54 \\
S2 & 84.25 & 85.53 & 82.25 & 84.01  \\
S3 & 82.56 & 83.27 & 82.50 & 82.77  \\
\hline \hline
\end{tabular}}
\end{table}

\begin{figure*}[bp]
\subfigure[]{
\begin{minipage}{0.5\linewidth}
\centering%
\includegraphics[scale=0.55]{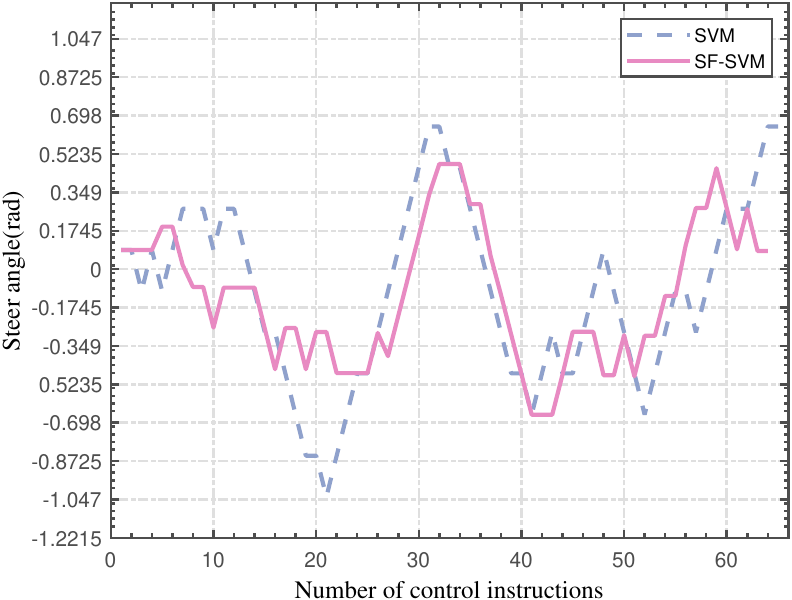} \label{fig3-4a}
\end{minipage}}
\subfigure[]{
\begin{minipage}{0.5\linewidth}
\centering%
\includegraphics[scale=0.55]{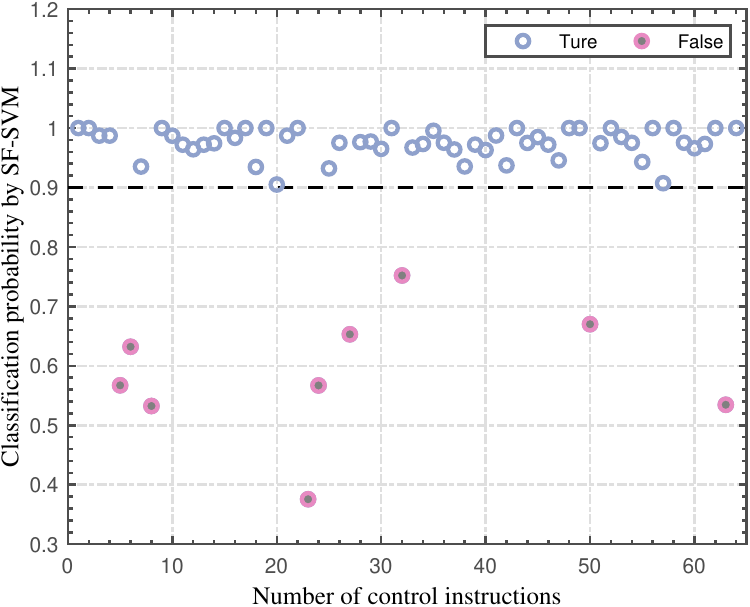} \label{fig3-4b}
\end{minipage}}
\caption{Input steering angle and classification probability in the scene 1: (a) input steering angle, (b) classification probability results with SF-SVM.}
\end{figure*}

\begin{figure*}[bp]
\subfigure[]{
\begin{minipage}{0.5\linewidth}
\centering%
\includegraphics[scale=0.55]{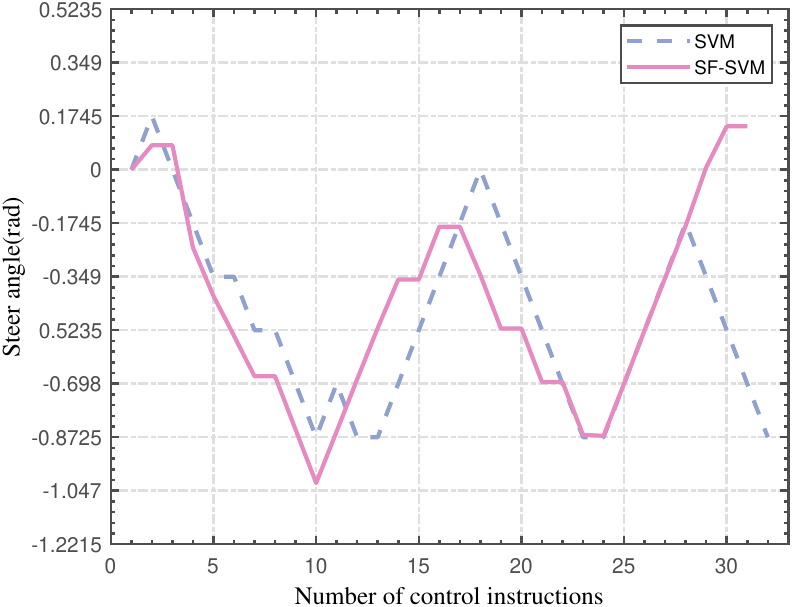} \label{fig3-6a}
\end{minipage}}
\subfigure[]{
\begin{minipage}{0.5\linewidth}
\centering%
\includegraphics[scale=0.55]{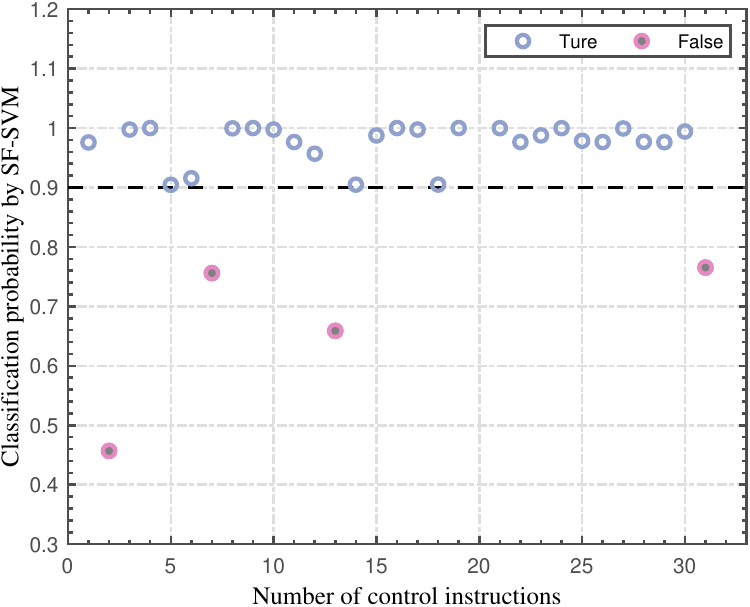} \label{fig3-6b}
\end{minipage}}
\caption{Input steering angle and classification probability in the scene 2: (a) Input steering angle, (b) classification probability results with SF-SVM.}
\end{figure*}

\begin{figure}[tp]
\subfigure[]{
\begin{minipage}{1\linewidth}
\centering%
\includegraphics[scale=0.5]{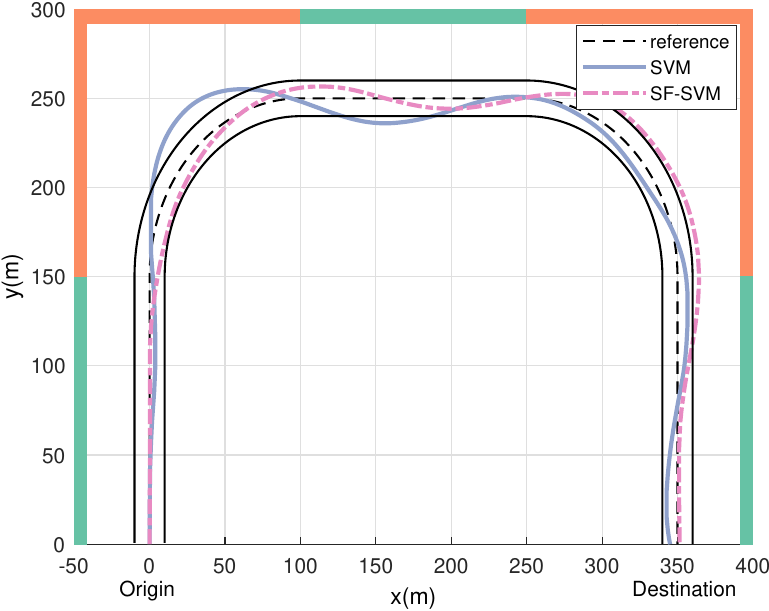} \label{fig3-3a}
\end{minipage}}
\subfigure[]{
\begin{minipage}{1\linewidth}
\centering%
\includegraphics[scale=0.5]{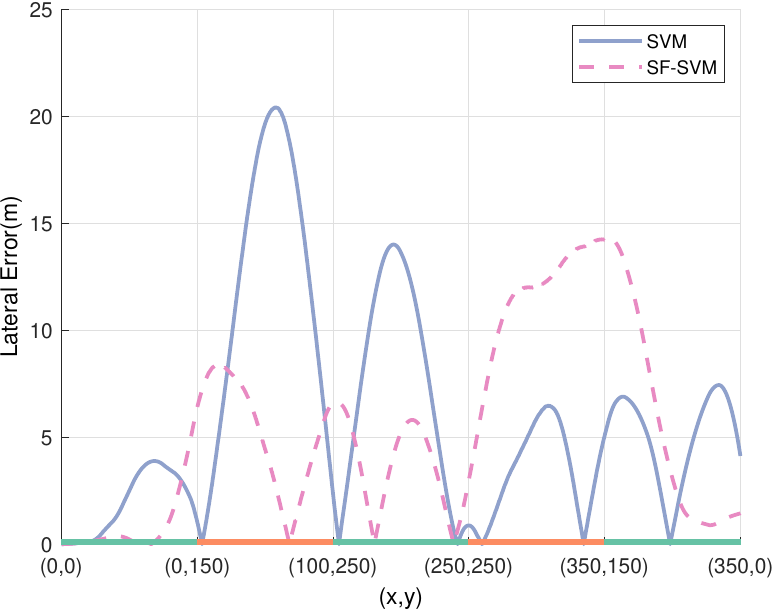} \label{fig3-3b}
\end{minipage}}
\subfigure[]{
\begin{minipage}{1\linewidth}
\centering%
\includegraphics[scale=0.5]{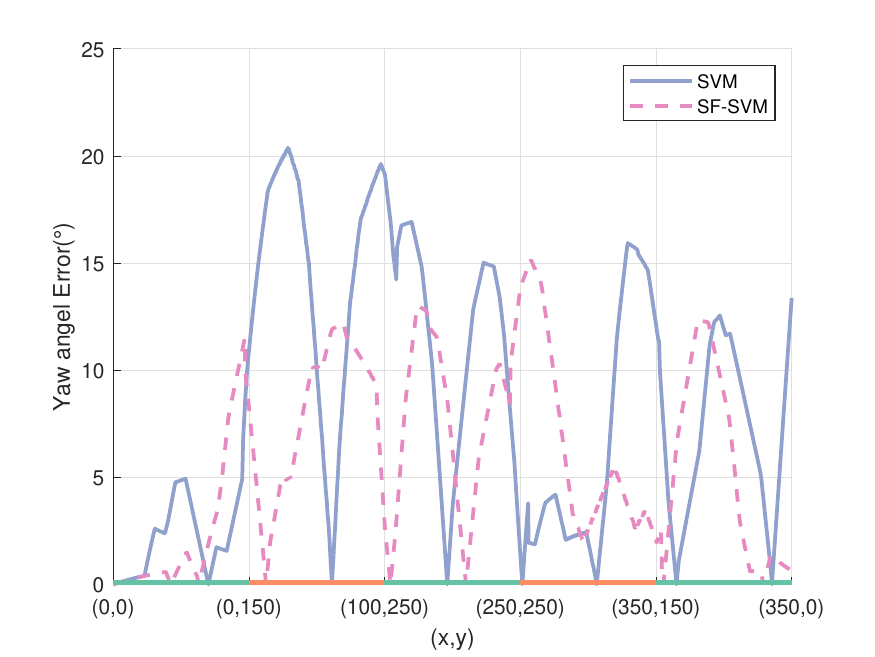} \label{fig3-3c}
\end{minipage}}
\subfigure[]{
\begin{minipage}{1\linewidth}
\centering%
\includegraphics[scale=0.5]{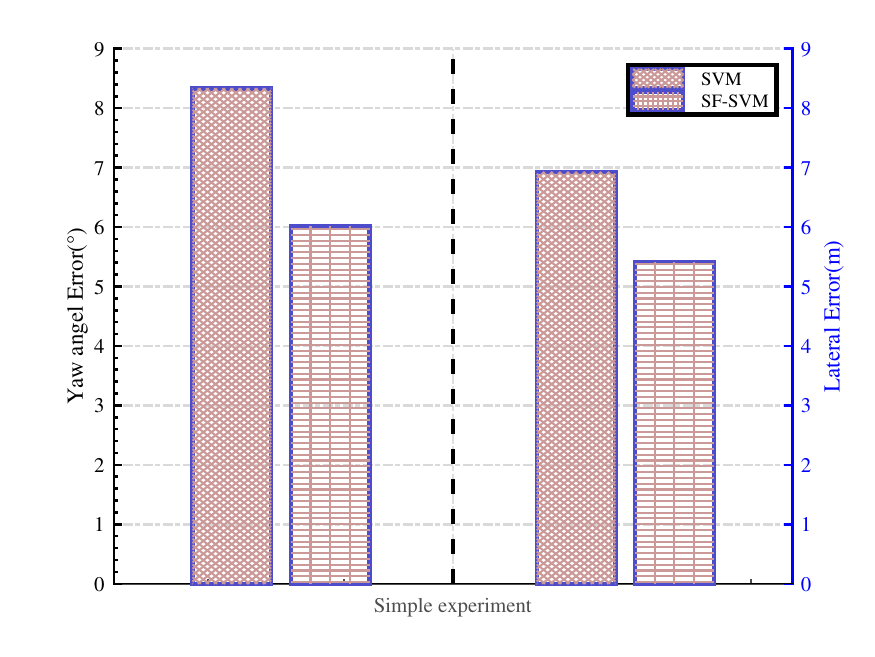} \label{fig3-3d}
\end{minipage}}
\caption{Road-keeping experiment with the SVM and the SF-SVM for subject one in the scene 1: (a) lateral displacement, (b) lateral error, (c) yaw angle error, (d) the mean value of lateral error and yaw angle.}
\end{figure}

\begin{figure}[tp]
\subfigure[]{
\begin{minipage}{1\linewidth}
\centering%
\includegraphics[scale=0.5]{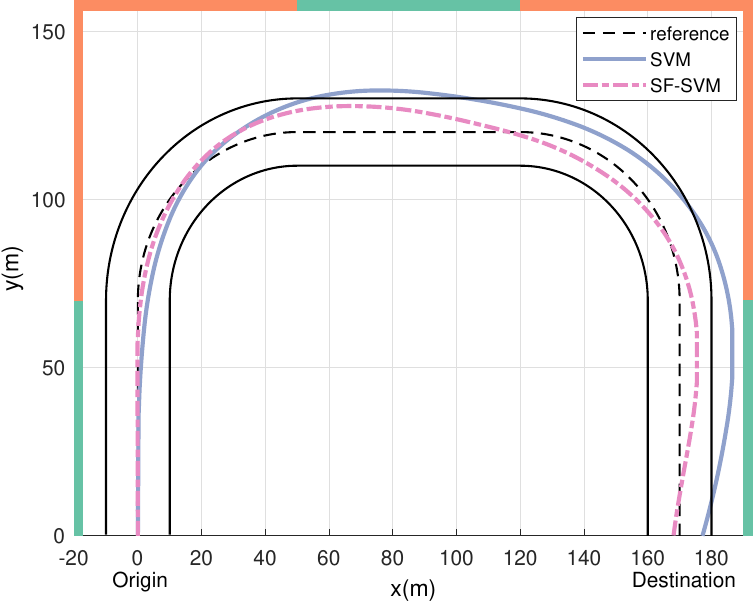} \label{fig3-5a}
\end{minipage}}
\subfigure[]{
\begin{minipage}{1\linewidth}
\centering%
\includegraphics[scale=0.50]{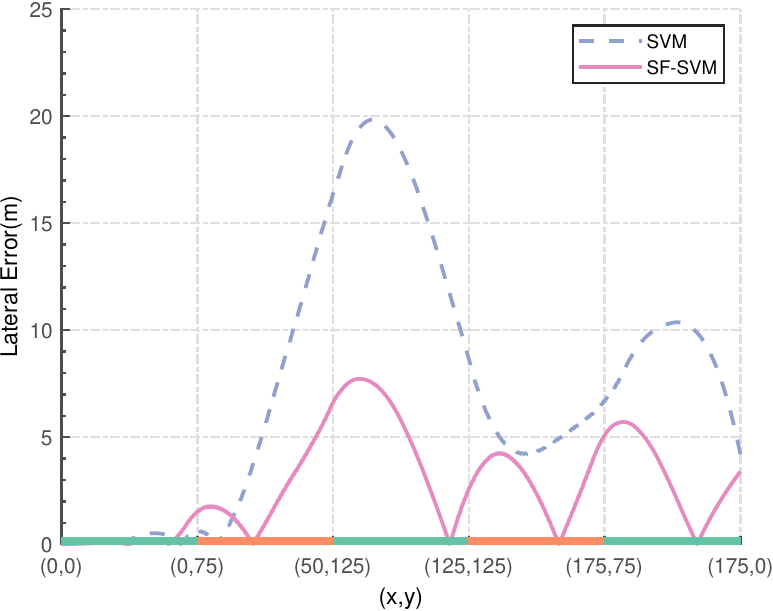} \label{fig3-5b}
\end{minipage}}
\subfigure[]{
\begin{minipage}{1\linewidth}
\centering%
\includegraphics[scale=0.50]{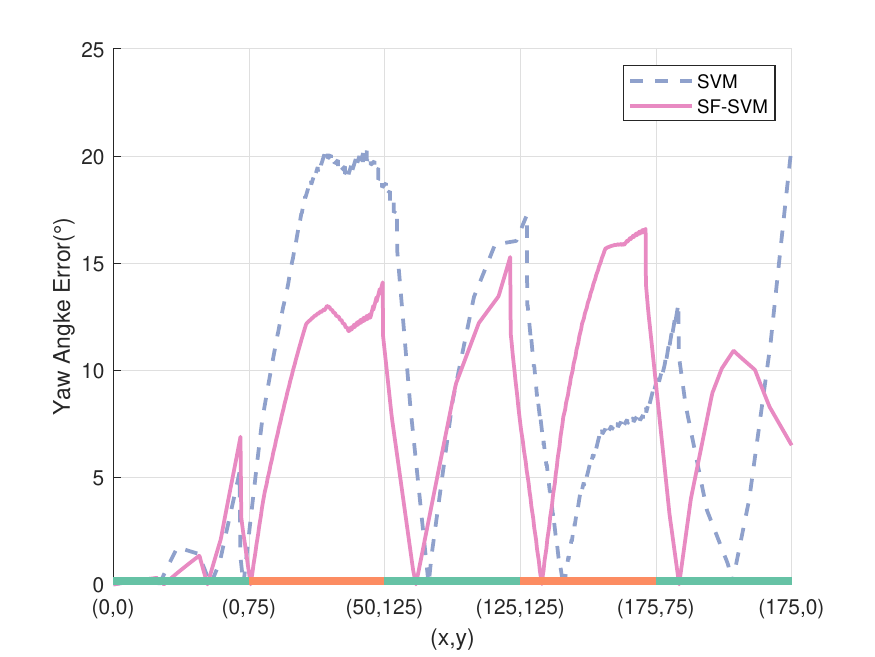} \label{fig3-5c}
\end{minipage}}
\subfigure[]{
\begin{minipage}{1\linewidth}
\centering%
\includegraphics[scale=0.5]{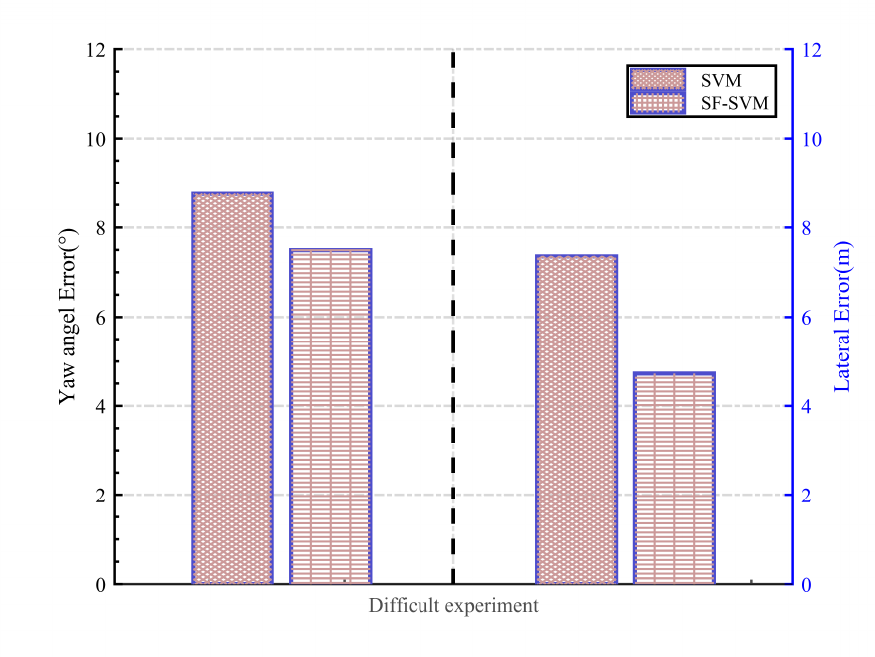} \label{fig3-5d}
\end{minipage}}
\caption{Road-keeping experiment with the SVM and the SF-SVM for subject one in the scene 2: (a) lateral displacement, (b) lateral error, (c) yaw angle error, (d) the mean value of lateral error and yaw angle.}
\end{figure}
\section{Road-Keeping Experiment} \label{sec4}
The road-keeping experiments are conducted separately in the scenes 1 and 2. The task scenes are shown in Figs.\ref{fig3-1}. The road parameters are shown in Table \ref{tab2}. The goal of the road-keeping test is to keep the vehicle running along the road centerline.
\begin{table}
\caption{Comparison of the parameters of both scenes} \label{tab2}
\centering
\renewcommand\arraystretch{1.2}
\setlength{\tabcolsep}{0.8mm}{
\begin{tabular}{cccc}
\hline \hline
Scene &  Radius(m) & \makecell[c]{Longitudinal speed(m/s)}& \makecell[c]{Straight length(m)}\\
\hline 1 & 100 & 2&150  \\
2 & 50 & 3 &70\\
\hline \hline
\end{tabular}}
\end{table}
\subsection{Experimental Process}
In the road-keeping experiments, the subjects are asked to conduct fifteen trials of road-keeping experiments in two class scenes with the SVM and the SF-SVM, respectively. The characteristics of the road in the scene 1 is long distance and small curvature, so the driving difficulty is low, but the average completion distance is large. The characteristics of the road in the scene 2 is short distance and the curve is large, so the driving task is more difficult, but the average completion time is short.

We apply the task completion rate $R$, the lateral error $L$ and the yaw angle $Y$ as metrics to evaluate the performance of the BCVs. A trial is considered successful if the task completion distance $D$ is shorter than the limit set to be two times the nominal distance \cite{zheng2014}. $L$ is defined as the distance from the vehicle center of gravity to the road centerline, and $Y$ is defined as the orientation error of the vehicle frame relative to the road. In this experiment, $\Delta\delta=10^{\circ}$, $\delta_ {\max}= 100^{\circ}$, $\delta_ {\min}=- 100^{\circ}$, the vehicle rotation ratio is 19. Subject one selects a time window length of 4s in the scene 1 and 3s in the scene 2.

\begin{remark}
The selection of the time window length is a compromise after taking accuracy and ITR into account. Generally, we will choose the length of the time window with the highest ITR, but more must be considered in practice. The road-keeping experiment in the scene 1, it takes more time to complete due to the long distance. As a result, subjects needs to send more classification instructions and error instructions will also increase. Therefore, subject one choose to use the longest time window length to improve the accuracy of the experiment and ensure the completion of the experiment.
\end{remark}

\subsection{Experiment Results and Analysis}
This section shows the experimental results for subject one. We choose one of trials to display the results. The green rectangle in figures represent the straight section, and the orange represents the turning section. For convenience, we have made the following notations: maximum lateral error $L_\textrm{m}$, mean lateral error $L_\textrm{a}$, maximum yaw angle $Y_\textrm{m}$, mean yaw angle $Y_\textrm{a}$.

\subsubsection{Experiment results in the Scene 1}
Subject one successfully completed 11 trials and 7 trials with SF-SVM and SVM, respectively. Fig.\ref{fig3-4a} shows input steer angle can only be an integer multiple of $\Delta \delta$ with the SVM. By contract, with the SF-SVM, the control module can flexibly output various steer angle. Fig.\ref{fig3-4b} shows the output probability value with the SF-SVM, and there are 65 instructions in this trial, with 9 probability values below 0.90. It can be seen that the online classification accuracy of this trial is 86.15\%.

Fig.\ref{fig3-3a}, \ref{fig3-3b}, \ref{fig3-3c} show the position, $L$ and $Y$ of the brain vehicle, respectively. Fig.\ref{fig3-3d} shows the $L_\textrm{a}$ and $Y_\textrm{a}$. As shown in Table \ref{tab4}, compared with the SVM, the SF-SVM generates the smaller average of lateral error and yaw angle error, the smaller maximum of lateral error and yaw angle error and the shorter task competition distance. Specifically, the task completion rate increased by 25.6\%, the average of lateral error is reduced by 10.15\%, and the average of yaw angle error is reduced by 16.88\%.

\begin{table}
\centering
\caption{Performance Metrics Comparison of the Road-Keeping experiment for Subject One in the scene 1.} \label{tab4}
\renewcommand\arraystretch{2}
\setlength{\tabcolsep}{1mm}{
\begin{tabular}  {c|cccccccc}
\hline \hline
\diagbox{Method}{Metrics} & $R$(\%) & $D$(m)& $L_\textrm{a}$(m) & $Y_\textrm{a}$(deg) & $L_\textrm{m}$(m) &$Y_\textrm{m}$(deg) \\
\hline with SVM & 46.67 &792.2707 & 6.0226 & 8.3320 &20.4132 &20.3878 \\
with SF-SVM & 73.33 & 780.2104 & 5.4116 & 6.9235 & 14.2673 &15.16\\
\hline \hline
\end{tabular}}
\end{table}

\subsubsection{Experiment results in the Scene 2}
Subject one successfully completed 9 trials and 6 trials with SF-SVM and SVM, respectively. Fig.\ref{fig3-6a} shows the input angle. Fig.\ref{fig3-6b} shows the output probability value with the SF-SVM. There are 32 instructions in this trial, with 4 probability values below 0.90. It can be seen that the online classification accuracy is 87.50\%.

Fig.\ref{fig3-5a}, \ref{fig3-5b}, \ref{fig3-5c} show the position, $L$ and $Y$ of the brain vehicle, respectively. Fig.\ref{fig3-5d} shows the $L_\textrm{a}$ and $Y_\textrm{a}$. As shown in Table \ref{tab5}, the SF-SVM model produces better performance. Specifically, the task completion rate increased by 20.0\%, the average of lateral error is reduced by 36.90\%, and the average of yaw angle error is reduced by 15.95\%.

\begin{table}
\centering
\caption{Performance Metrics Comparison of the Road-Keeping experiment for Subject One in the Scene 2} \label{tab5}
\renewcommand\arraystretch{2}
\setlength{\tabcolsep}{1mm}{
\begin{tabular}  {c|cccccccc}
\hline \hline
\diagbox{Method}{Metrics}  & $R$(\%) & $D$(m)& $L_\textrm{a}$(m) & $Y_\textrm{a}$(deg) & $L_\textrm{m}$(m) &$Y_\textrm{m}$(deg) \\
\hline with SVM & 40.00 &396.2591 & 7.4912 & 8.7471 &19.8352 &20.4915 \\
with SF-SVM & 60.00 & 384.2536 & 4.7271 & 7.3521 & 7.7234 &16.5996 \\
\hline \hline
\end{tabular}}
\end{table}

\subsubsection{Further Disscusion}
In the scene 1, even if with the SF-SVM control module, the lateral error is large in the range of coordinates (250,250)-(350,150). Meanwhile, in the scene 2, the overall lateral error is low, but the yaw angle error is large. It should be noted that the SF-SVM lateral control module only can weaken the wrong commands for the whole driving and play a certain corrective role, but it cannot avoid wrong commands. For incorrect instructions, SF-SVM requires less correct instructions to correct them. Hence, before applying to BCV in the real environment, the key technique is to improve the performance of the brain instructions generation module
(including higher precision of signal acquisition, the increase in the number of output commands, etc.).

\section{Conclusions} \label{sec5}
This paper has proposed a vehicle lateral control module based on the SF-SVM to improve the BCVs' driving performance. The FBCSP algorithm is introduced in the brain instructions generation moudle which improves the SNR of EEG signals and ensures better decoding performance. In the case of poor performance of driving BCVs, the SF-SVM based on the sigmod-fitting method is trained. Based on this model, the control module with the SF-SVM is developed, which it not only can make the control instructions more flexible, but also can weaken the instructions in the case of incorrect classification. What's more, experiments in the two road-keeping scenes are conducted, and results show that the control module with SF-SVM improves the driving performance of BCVs.
This study provides some insights on how to enhance the online implementation performance of the brain control dynamic system.

\vspace{11pt}

\bibliographystyle{IEEEtran}
\bibliography{Reference}

\end{document}